# Evolution of Level Sets in Hamiltonian Parameter Space

**Fariel Shafee**
Department of Physics
Princeton University
Princeton, NJ 08540

*Abstract:*

We develop a formalism to study the use of the Level Set Method in the investigation of the evolution of observables in terms of parameters of the Hamiltonian, both of the system itself and of a control part. A simple example with an analytic solution available perturbatively is examined. We show that B-splines can quite accurately and smoothly interpolate surfaces corresponding to constant expectation value of observables which form the level sets as projections on a finite mesh of data. Lastly we make a brief preliminary scrutiny of the possibility of using temperature as a relevant parameter in ensembles of quantum systems.

# 1. INTRODUCTION

We have previously reported [1,2] on the applicability of the Level Set Method (LSM) [3-5] to quantum control problems. In that work we concentrated on the level sets in the state function space. The expectation value of an observable for a given system with fixed parameters of course depends on the superposition of the eigenstates of the operator corresponding to the observable and we studied how level sets with constant values of the observables could be defined in the Hilbert space and how control operators created from a control Hamiltonian could move the system in the Hilbert space in such a way as to preserve the expectation value of the observable (EVO). The path from a given state to another with a different EVO is of course not unique, and we commented, if laser pulses were chosen for control, as is possible in many current chemical contexts, how simple combinations of operators could attain the target.

In a different scenario the space to be addressed may not be the Hilbert space of state vectors, but that of the parameters determining the potential and hence the Hamiltonian. For example, if we are interested in the ground state energy, or the energy of the first excited state, or in a specific spectral line, then we are not concerned with the mixture of states in determining the observable in issue, and therefore in the orientation of the state vector in Hilbert space. On the other hand virtually all observables, even those belonging to pure states, or to specific combinations of eigenstates will in general depend on the parameters of the Hamiltonian and when more than one parameter is involved, we expect to get contours in the parameter space which may be attacked by techniques developed by the LSM in a classical context.

In the next section we shall first try to make these notions mathematically more precise in developing a framework for the LSM for the space of Hamiltonian parameters. In the following section we shall study a few simple examples with analytic solutions. In section 4 we shall be concerned with numerical solutions in more complicated cases where this method may actually find appropriate use. In the last section we present our conclusions regarding possibilities and shortcomings of this approach.

# 2. LEVEL SETS IN HAMILTONIAN SPACE

Let us consider the system Hamiltonian

$$H_S(\mathrm{r},\mathrm{p},a_I) = \frac{\mathrm{p}^2}{2m} + V(a_i,\mathrm{r}) \qquad (1)$$

where the $a_i$ are parameters of the potential.

Obviously all eigenstates of the Hamiltonian will in general be functions of the $a_i$, and hence any expectation value

$$<\theta> = <\psi(a_i)|\theta|\psi(a_i)> \qquad (2)$$

will also be a function of the $a_i$.

Hence in the $a_i$ space we expect to have the equal EVO contours

$$\theta(a_i) = c \qquad (3)$$

If now we also introduce a control Hamiltonian interacting with the system

$$H_c = H_c(b_j) \qquad (4)$$

with the parameters $b_j$, then the equal EVO contours will be represented in the product space of the two sets of parameters, because the eigenstates will be perturbed to new functional forms involving the $b_j$ parameters:

$$\theta(a_i, b_j) = c \qquad (5)$$

It is possible that one or more of the parameters are environmental ones, affecting both $H_s$ and $H_c$. Let us call this common set $e_k$. This would reduce the dimensionality of the parameter space, but not change anything else. Let us choose one of these common parameters and call it $s$. Then

$$\theta(a, b, s) = c \qquad (6)$$

where we have dropped the indices from the noncommon parameters $a$ and $b$. If we have only one $a$ and one $b$, then in three dimensions Eq. 6 corresponds to a surface, and for a particular s, we get the intersection of the surface with the relevant plane giving us a contour. As $s$ changes the contour also changes, and we have a "motion" of the level set, such that each point *(a,b)* on the contour moves to a new point *(a',b')* on a different contour in the parameter space with changing $s$ (Fig. 1, 2).

Hence we can talk about the points on the level set having a velocity in the parameter space with $s$ representing time.

$$\begin{aligned} u_a &= \tfrac{da}{ds} \\ u_b &= \tfrac{db}{ds} \end{aligned} \qquad (7)$$

Since for any point the co-ordinates *(a,b)* in the parameter space is a function of *s*     a and b are functions of *s,* and differentiating Eq. 6 we get

$$\partial_s \theta + \vec{u}.\nabla_{a,b}\,\theta = 0 \qquad (8)$$

which is exactly analogous to the level set equation of motion for an interface [3].

A pertinent question here may be whether we can associate any particular point on one level set with another point on another level set. In the case of fluid motion we have no problem, in theory, in placing markers and identifying the motion of individual particles, so that **u** has an unambiguous meaning. In the present context of quantum Hamiltonians the points on the level sets cannot be so marked and identified. So it is possible that the "flow" of the level set is not normal to the contour curve (or surface, in higher dimensions) and points on one contour can spiral into another (like cyclonic and anticyclonic motion, with a tangential component of speed caused by a sort of Coriolis acceleration ( Fig. 3). Indeed it is even possible that the same curve can be a space-filling one and cover a two–dimensional surface area, i.e. beginning at a particular point $(a_o, b_o)$ in parameter space, we can reach any other point $(a,b)$ in a two dimensional space in the neighborhood of this point by varying *s*.

However, we should also remember that our potentials and Hamiltonians, both the system and the control part, will in general be analytic functions of the parameters, and self-avoiding space-filling curves such as the Peano curve (3) are not analytic functions of the parameters at any point. Non-self-avoiding curves also unacceptable as level sets in the quantum context, because bifurcations also reflect non-analyticity, though possibly of a milder type, and at a countable number of points.

If for simplicity we are only interested in the evolution of the level set and not of individual points on the level set, we may use the analogy with interface motion as the simplest possible algorithmic scheme. We, therefore, shall assume in the following that the individual points on the level sets move normally to the contour curves.

In the case of classical motion, one introduces a normal speed function, usually from empirical data or from a simple model, and also a curvature dependent force to smooth out irregularities and cusps and acnodes, which are usually eliminated by physical curvature dependent dynamical features of the interface, such as viscosity.

In the present quantum case, if we know the theory completely and can solve exactly for the EVO in terms of the parameters for all s, a and b, then the LSM is of course redundant. In reality quantum systems are almost never analytically fully soluble. We can then measure the observable at discrete values $(a_i, b_i)$ of the parameters at any given s on a grid and find the level set as per the algorithm used in interface motion (3).

The algorithm for the determination of the level set at a fixed value of *s* consists of
1) determining the EOV at the outermost grid points first;
2) checking the value at each grid point with the neighboring points, in both the horizontal and vertical directions;
3) continuing this procedure until we come across a grid point such that one of the neighbors has a value lower than c of Eq. (6)  (or it may be the other way around if the outermost points have EOV < c;
4) marking these boundary points as members of the level set.

This level set is a collection of neighboring grid points, and if the density of grid points is sufficient, it may be in practice indistinguishable from an analytic curve, i.e. we need the lattice spacing smaller than the experimental resolution or error.

We can then proceed to do the same for different *s* values and form a wire frame representing a surface. Now finite difference methods would allow us to interpolate between the discrete grid points and we can find the *b* on the surface (Eq. 6) for any continuous values of *a, s* and *c*.

In other words if we know that the environmental variable *s* has changed to a new value and the system parameter *a* has also changed to a new value, then we can find the corresponding value of *b*, which would keep EVO  *c* constant.

The smoothness criteria can be included into the mesh interpolation algorithm by usual numerical techniques, such as the application of splines and/or least square best-fit approximations. This would be equivalent to a viscous force needed in the analogous classical case.

## 3. SIMPLE ANALYTIC EXAMPLE

Let us first consider the trivial case of a simple harmonic oscillator.

$$H(w) = \frac{p^2}{2m} + \tfrac{1}{2} m\omega^2 x^2 \tag{9}$$

which has just one parameter.

We can consider the dipole transition element as the observable. Though a harmonic oscillator has no static expectation value of a dipole moment due to parity conservation, the dipole transition matrix element

$$d_{01}(\omega) = <1(\omega)|x|0(\omega)> \tag{10}$$

can be measured from the induced transition rate from the ground state to the first excited state by electromagnetic waves, e.g. from a laser.

Let us now introduce a Hamiltonian with also an anharmonic term as the control:

$$H_c(\omega, \varepsilon, b) = \tfrac{1}{2}\varepsilon m\omega^2 x^2 + bx \tag{11}$$

where we assume that $\varepsilon$ and $b$ are both much smaller than $\omega$.

The $x^2$ part simply modifies the frequency to a new value

$$\omega' = \omega\sqrt{1+\varepsilon} \tag{12}$$

The anharmonic term will change the eigenstates and hence the dipole moment. We finally get with first order corrections:

$$\begin{aligned}d_{01}(\omega,\varepsilon,b) &= <1,\omega,\varepsilon,b|x|0,\omega,\varepsilon,b> = \\ &[1 - b^2\, d_{01}(\omega,0,0)/\Delta^2\,]d_{01}(\omega,0,0)(1+\varepsilon)/N\end{aligned} \tag{13}$$

where

$$\begin{aligned}\Delta &= E_1 - E_0 = \hbar\omega' \\ N &= [1 + b^2\, d_{01}(\omega,0,0)/\Delta^2\,]\end{aligned} \tag{14}$$

Such an anharmonic term may from a small uniform electric field.

We can show this $d_{01}$ level set in a plane with the control parameters $\varepsilon$ and $b$ as the axes.

So, if it is observed that an ensemble of systems have a drifting dipole transition rate for any reason, known or unknown, we can use the control parameters ε and b to restore the EVO.

## 4. NUMERICAL APPROACH

As we have remarked earlier, in most systems it is not possible to extract an analytic expression for the observable in terms of the system and control parameters and one must resort to numerics. In systems with some unknown parameters it may be necessary to obtain the level sets experimentally on a discrete grid from a finite number of measurements. One can then use finite difference methods to interpolate the level set contours at intermediate points within the cells of the grid and for continuous ω. This might give us the velocities of the level set in terms of the grid coordinates and ω that may be used to predict the progress of the level set at continuous values of ω. The goodness of such an approximation would depend on the smoothness of the dependence of the EVO on the parameters chosen.

An implicit surface equation such as Eq. (13) in general has formal equivalents by rotating the variables and the function. So the EVO can become an argument in:

$$b = b(\omega, \varepsilon, d)$$
$$\varepsilon = \varepsilon(\omega, b, d) \tag{15}$$

So that given any specific value of the EVO ($d$ in this case), and of some of the control parameters, we can find the value of the remaining control parameter which keeps the EVO constant from the corresponding point on the surface. However, the equivalence is only formal. The inversion of nonlinear equations in exact analytic forms is of course in general impossible, and that is what makes the problem tackled by the LSM nontrivial.

We here illustrate the method for the simple example we have solved analytically above. Let us suppose that by experimentation we have sufficient data to pick up the some combinations of the system and control parameters which give the right EVO within an acceptable range, either directly, or by simple interpolation. So we have a number of data points belonging to the required surface. Now we can use different numerical methods to approximate the unknown surface from this finite and possibly quite small set of data, which can then be used to predict the control parameters required to keep the EOV constant when the system parameter drifts. A good candidate for surface reconstruction from a finite set of points is the B-spline interpolation method. In this method the surface is represented by the parametric form ( x(u,v), y(u,v), z(u,v) ), where the coordinates are in the physical parameter space of the Hamiltonian and the orthogonal algebraic parameters u and v are not necessarily related directly to simple combinations of the physical parameters, but are needed to reconstruct the surface numerically. Knot values have to be calculated for u and v [6,7] and they define B-spline basis functions $N_{i,p}(u)$ and $N_{j,q}(v)$ which depend on the [ $u_i$, $v_j$] mesh created by the knots and are actually Bernstein polynomials of degrees p and q, which may be chosen different in the two directions if the nature of the data so suggest. The surface is defined by

$$\vec{r}(u,v) = \sum_i \sum_j N_{i,p}(u) N_{j,q}(v) \vec{r}_{ij} \tag{16}$$

where $r_{ij}$ are the control points for the surface obtained from the given data points by solving the linear sets of equations (15). We can generate the whole surface by varying the interpolating variables u and v within the predefined domain, usually [ 0,1]. Details may be found in ref [6,7].

Before we show the results for our physical Hamiltonian, which is relatively simple, as a test of the accuracy of the numerical method we show in Fig. 4 and Fig.5 the surface for

$$x^3 + y^3 + x^2y - x\,y^2 - z = 0 \qquad (17)$$

drawn respectively by a *Mathematica* command, which simply connects the points without interpolating, and by our B-spline method which gives a smooth interpolating surface. In a spline fit of course the basis functions are regionally limited and hence the surface is not infinitely differentiable, but even a $C^2$ continuity with continuous derivatives even at joining points and infinite derivability at all other points gives an enormously more satisfying method of getting the constant EOV surface. The intersection of this surface with different z planes would give smooth level surfaces which can be used with reasonable confidence in control experiments, because we should remember that even in perturbation theoretic calculations, we rarely go beyond the second order, i.e. degree 2 in the perturbation parameter, and the predictions usually quite well with experiment.

## 5. TEMPERATURE AND EVO

At finite temperature T the EVO will depend on T, because at thermal equilibrium states with different energies will be occupied with different probabilities and the EVO may depend on the energy of the state. This does not apply when we are considering an observable defined uniquely in terms of a given set of states, such as $d_{01}$ above. But we may be interested in the average dipole transition moment, not just that between the first two states. In this case we have another parameter in the problem:

$$d(T,\omega,\varepsilon,b) = <n+1,\omega,\varepsilon,b|x|n,\omega,\varepsilon,b> \exp[-E_n/kT]/Z \qquad (18)$$

where Z is the partition function. In principle the problem is still tractable. In fact, we may in such a case probably keep $\omega$ as a constant and let T take its place, so that T becomes the scale variable. The similarity between T and and time is well-known in physics. In many-body theory we have to use temperature dependent Green's functions with terms like exp[ –E/kT] coming in with exp[ - i H t], so that i/kT plays the role of a kind of extended time dimension. So the concept of finite temperature level sets need to be investigated more thoroughly. It is very unlikely that one can get analytic closed solutions in nontrivial cases and LSM may indeed be a useful tool in such a situation.

Further work is in progress to investigate these ideas.

## 6. CONCLUSIONS

We have seen that in the context of quantum control where the parameters of the system Hamiltonian may drift, causing the expectation value of an observable to change, it is possible to obtain information for corrective action by changing the parameters of a control Hamiltonian by using the ideas of the level set method which hitherto have most been used in fluid dynamics and other classical contexts. Unlike the classical case, of course we do no use a force equation, as the evolution of the system has to be described in terms of system and control parameters and not in terms of real space and time. However, smoothness is still a key issue and we have shown here that the B-spline method is an excellent tool to produce smooth interpolating surfaces. It is known that varieties of this method, using different techniques to parameterize the input data points to generate the surface (and hence the level sets) such as the equidistant method, equal chord method etc. can be useful for different types of data. In particular the equal chord method is seen to provide a tool for damping out high curvatures, which is achieved by hand in the usual fluid dynamic context as an additive to the apparent dynamical force.

The case for treating temperature as a relevant parameter for ensembles of systems at finite temperature probably deserves further study.

The author thanks Professor H. Rabitz for encouragement and discussions.

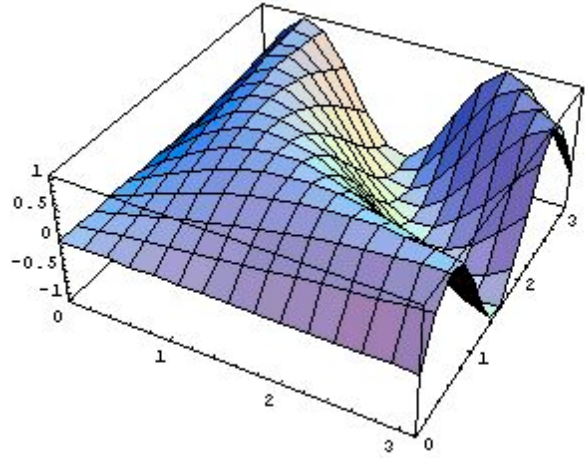

Fig.1 : Surface of   d (a,b,s) = s - sin ( ab) =0

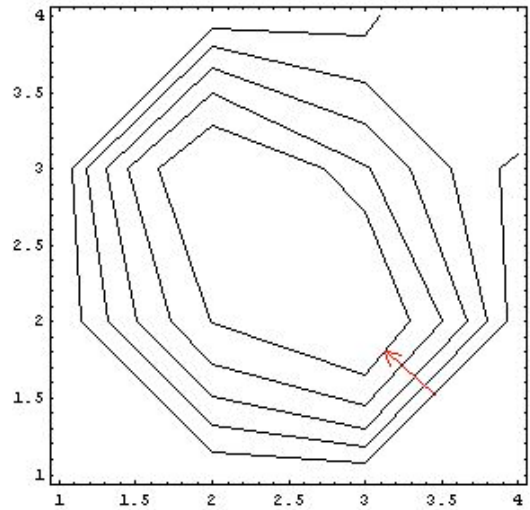

Fig. 2: Level set contours  d(a,b,s) =0 as above for different s values.
The red arrow shows normal velocity.

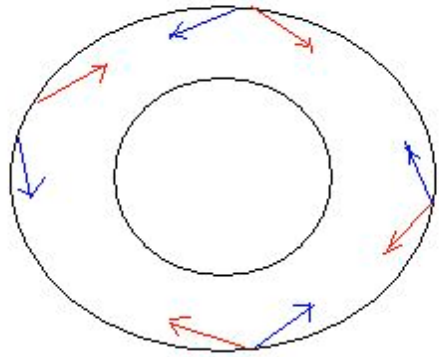

Fig. 3: Motion of level sets may not in general be normal, but may have negative or positive vorticity

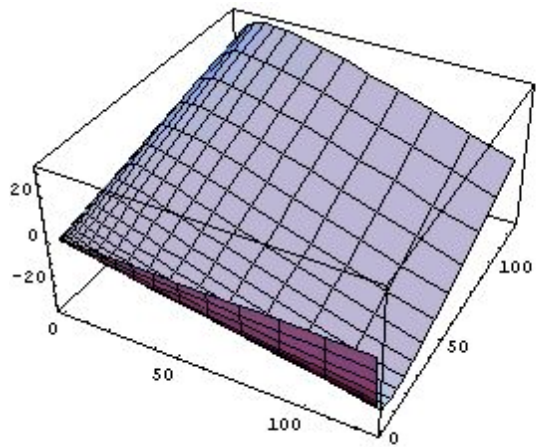

Fig. 4: Surface drawn by using the exact function $x^3 + y^3 + x^2y - xy^2 - z = 0$ (*Mathematica*)

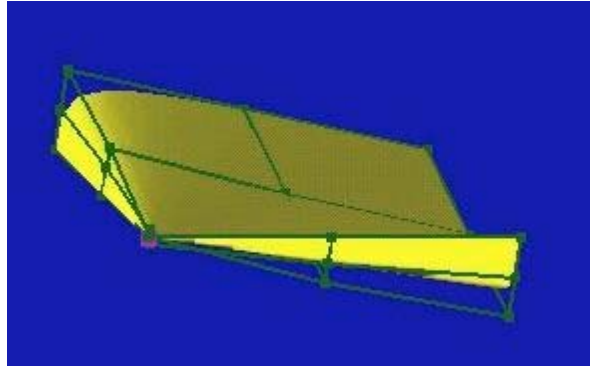

Fig 5: The same surface regenerated by using cubic splines ( i.e. p = q = 3) from 25 data points generated by the formula (seen from a slightly different perspective).

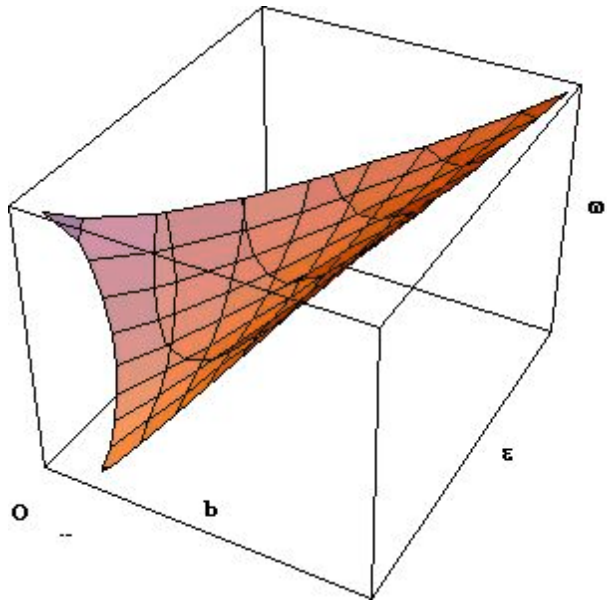

Fig 6: Plot of surface with equal EOV ( $d_{10}$= 1.0) , against b , $\varepsilon$ and $\omega$ along the three co-ordinate axes drawn from Eq. 13. O is the origin, and Hamiltonian parameters domains are 0<b<1, 0<$\varepsilon$<1, and 0.3 < $\omega$< 1.0.

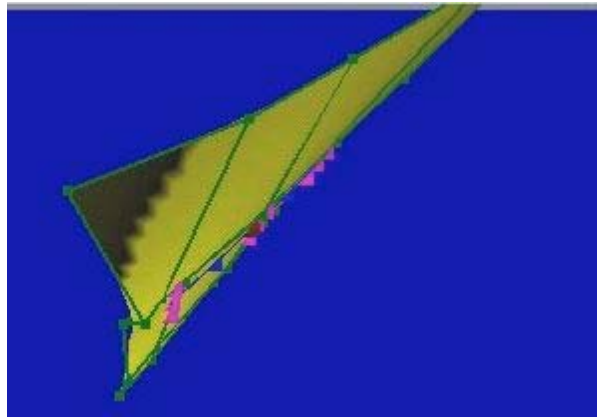

Fig 7: The same surface generated from an array of 25 "data points" generated from the formula.